\documentclass{aastex}
\usepackage{emulateapj5}

\def\etal{{et~al.}}
\def\amin{\ifmmode^{\prime}\else$^{\prime}$\fi}
\def\asec{\ifmmode^{\prime\prime}\else$^{\prime\prime}$\fi}

\def\simgt{\lower.5ex\hbox{$\; \buildrel > \over \sim \;$}}
\def\simlt{\lower.5ex\hbox{$\; \buildrel < \over \sim \;$}}

\newcommand\asca{{\it ASCA}}
\newcommand\chandra{{\it Chandra}}

\def\etal{{\it et~al.}}
\def\amin{\ifmmode^{\prime}\else$^{\prime}$\fi}
\def\asec{\ifmmode^{\prime\prime}\else$^{\prime\prime}$\fi}
\def\nhunits{\times 10^{22}\ {\rm cm^{-2}}}
\def\psr{\hbox{PSR~J1846$-$0258}}
\def\snr{\hbox{G29.7-0.3}}
\def\kes{\hbox{Kes~75}}

\slugcomment{Version of Sep 17 2002}

\shortauthors{Helfand, Collins, \& Gotthelf}
\shorttitle{Chandra X-ray Imaging Spectroscopy of SNR Kes~75}

\begin{document}

\title{Chandra X-ray Imaging Spectroscopy of the \\ Young Supernova Remnant Kesteven~75}

\author{David J. Helfand, Benjamin F. Collins, and E. V. Gotthelf}
\affil{Columbia Astrophysics Laboratory, Columbia University,
550 West 120$^{th}$ Street, New York, NY
10027, USA; djh@astro.columbia.edu; bfc@astro.columbia.edu; evg@astro.columbia.edu}

\begin{abstract}
We present a spatially resolved spectroscopic analysis of the young
Galactic supernova remnant \kes\ (SNR G29.7-0.3) using the \chandra\
X-ray Observatory.  \kes\ is one of an increasing number of examples
of a shell-type remnant with a central pulsar powering an extended
radio/X-ray core.  We are able to pinpoint the location of the
recently discovered pulsar, \psr, and confirm that X-rays from the
remnant's core component are consistent with non-thermal power-law emission from
both the pulsar and its surrounding wind nebula. We find that the
spectrum of the pulsar is significantly harder than that of the wind
nebula. Fainter, diffuse emission is detected from throughout the
volume delineated by the radio shell with a surface brightness
distribution strikingly similar to the radio emission. The presence of
strong lines attributable to ionized Mg, Si, and S indicate that at
least some of this emission is thermal in nature. However, when we
characterize the emission using a model of an underionized plasma with
non-solar elemental abundances, we find we require an additional
diffuse high-energy component. We show that a significant fraction of
this emission is an X-ray scattering halo from the pulsar and its wind
nebula, although a nonthermal contribution from electrons accelerated
in the shock cannot be excluded.

\end{abstract}

\keywords{pulsars: individual (\psr); supernova remnants:
individual(\snr); star: individual (\psr); stars: neutron;
X-rays: general}

\section {Introduction}

The expected products of a core-collapse supernova include a shock
wave propagating into the circumstellar/interstellar medium,
metal-enriched ejecta from the progenitor star moving outward behind
the shock, and a collapsed remnant of the stellar core (often a
rapidly spinning, magnetized neutron star). These products should
manifest themselves as a limb-brightened shell of radio emission at
the outward-moving shock, an interior volume of hot, shocked gas and
stellar ejecta emitting thermal X-rays, a short-period radio and/or
X-ray-$/\gamma$-ray pulsar, and a bright central nebula 
driven by the pulsar's relativistic outflow. To date, more
than 1300 pulsars (Manchester et al. 2002), 230 radio supernova remnant (SNRs -- Green 2001) (80 with
thermal X-ray emission), and two dozen pulsar wind nebulae
(PWNe -- Kaspi and Helfand 2002) are known in the Milky Way; the only problem is that they are
seldom all found in the same place. The rare coincidences that
actually match our expectations for the grave sites of massive stars
are called composite supernova remnants (Weiler 1983; Helfand and Becker 1987).

The Galactic radio source \kes\ (SNR \snr)
is a prototypical example of a composite remnant. It
exhibits a limb-brightened, $3\farcm5$-diameter radio shell with a
steep spectrum ($\alpha \sim-0.7$, where $S(\nu) = S(\nu_0)\nu^{\alpha}$) and modest radio polarization,
surrounding a flat-spectrum ($\alpha \sim0.0$), highly polarized radio
core first detected in X-rays by the Einstein Observatory (Becker,
Helfand \& Szymkowiak 1983). Neutral hydrogen absorption measurements
show the source to lie beyond the solar circle on the far side of the
Galaxy at a distance of $\sim19$~kpc (Becker \& Helfand 1984).
Observations by the Advanced Satellite for Cosmology and Astrophysics
(\asca) verified the existence of both thermal and nonthermal
emission, but lacked the spatial resolution necessary to separate the
components (Blanton \& Helfand 1996).  A recent observation with the
X-ray Timing Explorer led to the discovery of a young, energetic X-ray
pulsar, \psr, with a characteristic age of $723$ yrs.
The pulsar was localized to within $1^{\prime}$ of the remnant core
using an \asca\ observation (Gotthelf et al. 2000). Recent pulsar 
timing observations 
suggest that the true age of the remnant could be between 980 and 1770 years
(Mereghetti et al. 2002).

We report here the results of an observation of \kes\ with the
\chandra\ X-ray Observatory. We utilize the spatial resolution of
\chandra\ to determine an accurate location for \psr, finding it at
the center of a highly structured pulsar wind nebula. The spectral
properties of the pulsar, its nebula, and the surrounding shell are
explored, yielding evidence for a nonthermal component coincident with
the thermal shell X-ray emission.  In \S 2 we describe the
observations and data reduction procedures. In \S 3 we detail our
imaging results, while in \S4 and \S 5 we present our spectral results
on the core and shell components, respectively.  We conclude (\S 6)
with a discussion of these results and their implications for \kes\
itself, and for the class of composite remnants in general.

\section{Observations}

Our observation of \kes\ was obtained on $10-11$ Oct 2000 with the
\chandra\ X-ray Observatory (Weisskopf, O'Dell, \& van Speybroeck
1996). Photons were collected using the Advanced CCD Imaging
Spectrometer (ACIS), a mosaic of ten X-ray CCD chips, with the target
placed on the ACIS--S3 chip, offset $2^{\prime}$ from the nominal
aim-point to avoid losing portions of the remnant to the CCD gaps.
The back-side illuminated ACIS--S3 CCD is sensitive to photons in the
$\sim 0.2-10$ keV energy range with a spectral resolution of $E/\Delta
E \sim 10$ at 1 keV. The on-axis angular point-spread function (PSF)
of the telescope is $\sim 0\farcs5$ at 1 keV and is
undersampled by the CCD pixels ($\sim 0\farcs49 \times 0\farcs49$). As
a consequence of the 3~s CCD readout time, no timing information for
the pulsar ($P = 324$~ms) is available with the ACIS instrument.

Data reduction and analysis were performed using the CIAO, FTOOLS, and
XSPEC X-ray analysis software packages.  To correct for the
detrimental effects of charge transfer inefficiency (CTI) on the gain
and resolution of the CCD, we reprocessed the Level 1 event data
using the custom software of Townsley et al. (2000). The resulting
CTI-corrected Level 2 event file was then time-filtered to exclude
intervals of unacceptably high background activity caused by flares in
the rate of solar cosmic ray particles. These intervals
were identified using a $3\sigma$ iterative clipping algorithm applied
to a lightcurve created using all data on the ACIS-S3 chip but with
the supernova remnant emission excluded (defined as emission from a
radius of $r>100^{\prime\prime}$ centered on the SNR radio shell). The
subsequent time-filtering resulted in a useful exposure of 33.7 ks.
The instrument response files used in the spectral fits were
calculated using the RMF and quantum efficiency uniformity (QEU)
calibration files that accompany the CTI-removing software. 

\section{Image Analysis}

Figure 1 shows the \chandra\ X-ray view of the supernova remnant \kes.
As expected, we resolve a bright point-like source at the center of
the SNR shell, embedded in a compact X-ray nebula.  We take this as a
detection of \psr\ whose flux of 0.12 ct s$^{-1}$ accounts for
$\sim15\%$ of the $1-7$ keV emission from the core component (pulsar
$+$ compact nebula) of the remnant. Superposed on the X-ray image are
20 cm radio contours derived from data being taken for our new VLA
Galactic Plane survey (Helfand et al. 2001). The similarity between the two
images is striking: the core synchrotron nebula, the diffuse shell
emission, and the two limb-brightened areas in the southeast and
southwest quadrants of the shell are all co-spatial.

The astrometry given in our \chandra\ event file contains known
systematic errors\footnote{See
http://cxc.harvard.edu/cal/ASPECT}. To derive an
accurate position for \psr, we first corrected the astrometry
following the prescription recommended by the CXC\footnote{See
http://cxc.harvard.edu/cal/ASPECT/fix\_offset/fix\_offset.cgi}.  We
then ran the source-finding CIAO tool CELLDETECT with a
signal-to-noise threshold of 2.5 (employing LDETECT background
determinations) in order to locate sources with possible stellar
counterparts which could be used to determine the accuracy of the
corrected coordinates.  Comparing the resulting source positions to
the USNO astrometric catalog (Monet 1996) we found five coincidences.
One of these lies on top of the bright southwestern filament which
affected the X-ray source position determination; this was excluded
from further consideration. We then refined the \chandra\ positions by
spatially fitting each source with a model PSF using
SHERPA. We included a constant background level in this model to allow
for the bright nebular emission.  The weighted root-mean-squared
average of the offsets between our calculations and the coincident USNO
star positions is $0\farcs3$, of the same order as the typical uncertainty
of the USNO catalog (Deutsch 1999) and of the expected \chandra\
astrometric uncertainty. The computed offsets of $\Delta \rm{R.A.} =
-0\farcs04$ and $\Delta \rm{Dec} = 0\farcs09$ confirms the accuracy of
the corrected \chandra\ astrometry. The spatial fit for the position
of the pulsar yields J2000 coordinates $18^h46^m24.94^s\pm\ 0.01$, $-02^{\circ}
58^{\prime} 30\farcs1 \pm\ 0.2$. Comparison with the model PSF
confirms the detection of \psr\ as an unresolved point-source.

\section{Spectral Analysis}

It is clear from the image shown in Figure 1 that \psr\ is embedded
within a pulsar wind nebula. A complete spectral analysis of the pulsar
requires a model of the wind nebula to use as a background
component. It is further found to be necessary to account for the
effects of pile-up on the pulsar spectrum (to be discussed in \S
4.2). In the following, we first determined the absorbing column density from
spectral fits to the PWN whose spectrum in not distorted by
pile-up. In the subsequent fits to the pulsar spectrum we fix the
absorbing column to this value and use the PWN model as an additive
spectral component. Background subtraction for the pulsar wind nebula
itself was accomplished using a region southwest of the
nebula but within the SNR shell in order to account for any shell
emission underlying the nebula; for the remainder of the remnant,
background estimates were obtained from a source-free area on the
ACIS--S3 chip approximately $4^{\prime}$ northwest of the remnant;
errors in the count rates for all regions resulting from background
uncertainties are $\le2\%$.

For all spectral fitting presented herein the analysis is restrict to
the energy range of $1-7$ keV.  Below 1~keV, few source photons
are detected owing to the high absorption column density to the
remnant; above $7$ keV the spectrum falls off for all models.  Spectra
were grouped to contain a minimum of $20\ {\rm counts\ bin^{-1}}$ and
all errors on the spectral fits are quoted for a 1-$\sigma$ confidence
range. For spectra extracted from extended regions, it is no
longer sufficient to use an instrumental response appropriate for a
point source imaged at a single location on the detector. We have
therefore employed a user-supplied algorithm created by A. Vikhlinin\footnote{available at
asc.harvard.edu/cont-soft/software/calcrmf.1.08.html} 
for these regions to produce a count-weighted response function in
$16^{\prime\prime} \times16^{\prime\prime}$ bins, averaged over 
each extraction aperture.

\subsection{The Pulsar Wind Nebula}

Surrounding the pulsar is a highly structured diffuse nebula, $\sim
26^{\prime\prime} \ {\rm by} \sim 20^{\prime\prime}$ in extent, similar to
that seen in several other high-resolution \chandra\ images of pulsar
wind nebulae (Gotthelf 2001 and references therein). The morphology is
generally axisymmetric about a line $\sim30^{\circ}$ east of north,
with hot spots along this axis on either side of the pulsar. A blow-up
of the core region of \kes\ is show in Figure 2.

To derive the flux from the pulsar wind nebula, we define an
elliptical region with the pulsar region excluded (as shown in Figure
2a); this region contains a total of $\sim 23,000$
background-subtracted counts. Given the small extent of the nebula and
the fact that it lies within a single (FEF) response region, we used
the point source response matrix derived for the position of the
pulsar. The PWN spectrum is well characterized by an absorbed
power-law model which results in a spectral fit with
$\chi^2_{\nu}=0.97$ for 256 DoF. The best fit model parameters were
$\Gamma=1.92 \pm 0.04$ and $N_H = 3.96 \pm 0.08 \nhunits$.  The
unabsorbed flux in the $0.5-10$ keV band from the nebula is $4.0
\times 10^{-11}\ {\rm ergs\ cm^{-2}\ s^{-1}}$. The ACIS spectrum and
best fit model for the wind nebula region is shown in Figure 3.

To explore the possibility of spectral variations within the nebula,
we analyzed several morphologically distinct regions. These included a
2$\times$2 square grid of $17^{\prime\prime}$ cells centered on \psr,
two annular regions from $2\asec - 6\farcs4$ and $6\farcs4 - 15\asec$,
and several elliptical regions around distinct features of the wind
nebula including the northern and southern lobes and the knot-like
feature present in each (see Figure 2b for a definition of the latter
regions). The pulsar radiation was excluded in all cases by deleting
a $2^{\prime\prime}$-radius region around it. 
All regions are well-fit with a simple power-law plus
absorption model, and in each case the spectral parameters derived
agree to within $2\sigma$ both with each other and with the fit
derived for the nebula as a whole (e.g., see Figure 4). In particular,
there is no evidence for hardening of the spectrum with increasing
distance from the pulsar such as has been seen in G21.5-0.9 (Slane et
al. 2000) and 3C58 (Bocchino et al. 2001) and ascribed to synchrotron
losses. Table 1 summarizes these results.

\subsection{\psr}

A spectrum of the pulsar was obtained by extracting counts from within
a radius of $2^{\prime\prime}$ centered on the pulsar's position.  The
spectrum contains a total of 4004 counts (including only $\sim10$
background counts, but, assuming a nebular surface brightness equal to
that in the surrounding $2-4^{\prime\prime}$ annulus, a $\sim 15\%$
contamination from nebular background underlying the source region). A
naive fit to a simple power-law model including the effects of
interstellar absorption adequately describes the data, yielding a
reduced $\chi^2=0.9$ for 50 degrees of freedom (DoF).  The spectral
index of 1.02 $\pm0.10$ is somewhat flatter than that of the Crab
pulsar, but is similar to recently derived indices for other young
pulsars including Vela (Pavlov et al. 2001), PSR~J2229+6122 (Halpern
et al. 2001), and PSR~0540-69 in the LMC (Wang and Gotthelf
2000). However, the estimated pile-up fraction based on the count rate
is of order $\sim 10$\%, an effect that tends to flatten the
observed spectral index. The inferred column density is $N_H =
3.2\pm0.2 \times 10^{22}$ cm$^{-2}$ although, again, any artificial
flattening of the spectrum resulting from pile-up will likely bias
$N_H$ to lower values given the tight correlation between the two
parameters (see Figure 4).

Our ultimate model for the pulsar spectrum consists of four
components: interstellar absorption, CCD pile-up, a power-law
describing the pulsar, and another power-law describing the nebular
background.  We fixed the column density to 3.96 $\times 10^{22} \
{\rm cm^{-2}}$ and the slope of the background power-law to 1.92, both
results of fitting an absorbed power-law to the wind nebula emission.
From our estimate above, we fixed the normalization of the background
power-law to 15\% of the normalization of the source power-law
(varying this fraction from 10\% to 20\% had no significant effect on
the pulsar's spectral index). Fitting for the slope and normalization
of the source power-law as well as the grade-morphing parameter of the
pile-up model\footnote{See
http://space.mit.edu/\%7Edavis/papers/pileup2001.ps}, we performed a
thorough search of the parameter space to find the global minimum. The
result is a somewhat steeper spectral index for the pulsar emission of
$\Gamma= 1.39 \pm 0.04$. This fit implies a pile up fraction of $8\%$,
consistent with the expected value, and a corrected nonthermal flux
from the pulsar of $9.50 \times 10^{-12} \ {\rm ergs} \ {\rm cm}^{-2}
\ {\rm s}^{-1}$ in the $0.5 - 10$ keV band. The spectrum of both the
pulsar and the wind nebula are displayed in Figure 3, along with
residuals from the best fit models detailed in Table 1. Based on the
nebula's power-law index, the pulsar's spectral index is consistent
with that predicted by the empirical relationship of Gotthelf and
Olbert (2001) and lies near the mean of these quantities derived for
similar young pulsars with wind nebula (see Table 1 of Gotthelf and
Olbert 2001).

We also looked for low-energy excess X-ray flux which could represent
surface thermal emission from the young neutron star (see the
residuals for the four lowest energy spectral points in Figure
5). Although the number of counts present does not allow for a
detailed analysis, we fitted for a power-law index restricted to the
$2-7$ keV range where any thermal surface emission is expected to be
negligible. This yields $\Gamma = 1.31 \pm 0.06$ (Figure 5). Holding
this value (and the nebular $N_H$) fixed provides an estimate for the
blackbody temperature of $kT = 0.16 \pm 0.03$ keV, close to the
expected temperature of $1.8 \times 10^6$K for standard cooling of a
$\sim 1,000$-year-old neutron star (e.g., Umeda et al. 1994). However,
the error on the normalization is sufficiently large that no
meaningful measure of the emitting area is possible.

\section{Emission from the Supernova Remnant Shell}

In order to examine the spectral properties of the shell emission, we
have again defined several areas of interest (see Figure 6). We
considered i) the entire shell, defined as a circular aperture with a
radius of $100\asec$ centered on \psr\ and excluding the elliptical
region containing the pulsar and its nebula, ii) an ellipse around
each of the `clumps' in the south, and iii) three annuli defined so
as to correspond roughly with the boundaries of these (now excluded)
ellipses.

As a starting point from which to analyze the thermal shell, we
extracted a spectrum of everything outside the pulsar wind nebula to a
radius of $\sim100\asec$.  We fitted a model consisting of
interstellar absorption, a thermal bremsstrahlung continuum, and three
Gaussians representing the strongest emission lines, in order to
compare our results with a similar fit applied to the \asca\ data by
Blanton \& Helfand (1996).  The positions of the three Gaussians are
all consistent with the \asca\ analysis, and correspond to the ions
\ion{Mg}{11}, \ion{Si}{13}, and \ion{S}{15}.  The best-fit emission
temperature, $kT = 2.99 \pm 0.12\ {\rm keV}$, is significantly
discrepant with the \asca\ and BeppoSAX (Mereghetti et al. 2002)
values, however, a consequence of the inability of the latter observations to
distinguish spatially the various remnant components (see below). The
FWHM of the Si line offers a crude estimate of the remnant expansion
velocity of $\sim 3700$ km s$^{-1}$.

The heuristic model is also useful in searching for spatial variations
in the shell spectrum. We constructed fits for the two bright features
separately, for the diffuse emission outside these regions, and for
three annuli, as well as for the entire remnant (in all cases
excluding the synchrotron core). The results are displayed in Table
2. In all cases, the derived $N_H$ is significantly below that derived
from the power-law fit to the pulsar wind nebula; in some cases it is,
unphysically, less than the measured neutral hydrogen column density
(\S5), indicating an excess in soft flux over that accounted for by
the bremsstrahlung continuum. The derived temperatures suggest a
gradient, decreasing from the inside out, also suggestive of a problem
with the model.  The line centroids are largely constant throughout.
Unconstrained models of a Raymond-Smith thermal plasma (Raymond \&
Smith 1977), as well as multi-temperature plane-parallel shock, and
non-equilibrium ionization models (Hamilton \etal\ 1983; Borkowski
\etal\ 1994; Liedahl \etal\ 1995; Borkowski \etal\ 2000), including
interstellar absorption and varying abundances, all do a poor job of
fitting the spectrum.

The problem common to all of the thermal fits is that, in order to
account for the observed spectrum at higher energies, the continuum
temperature is forced to relatively high values inconsistent with the
line strengths, line centroids, and continuum shape below 2 keV
(Figure 7). For the \asca\ and BeppoSAX data in which the synchrotron
nebula was not resolved from the shell emission, this extra hard flux
was simply absorbed into the power-law component. By isolating and
excluding the central nebula's emission, the problem with the shell
spectrum becomes apparent.

In order to minimize the number of free parameters, we fixed the
column density to the well-determined value from the wind nebula fit:
$N_H = 3.96\nhunits$.  We then fitted a non-equilibrium ionization
model plus a power-law to account for the high energy flux (Figure
8). The abundances for H, He, C, N, O, Fe, and Ni for which we have no
usefully constraining data were held fixed at the solar value, while
the ions of Mg, Si, S, and Ca were allowed to vary independently. This
produces an acceptable fit ($\chi^2 = 2.15$ for 116 DoF) for the
diffuse shell emission excluding the two bright regions in the
south. The derived temperature is $\sim0.7$ keV, the power-law index
is $\sim 1.7$, and the ion abundances are $\sim 0.5$ solar. The fit to
the two clump regions combined (their spectra are statistically
indistinguishable) is somewhat worse; the derived temperature is very
similar, the abundances are slightly higher, and the power-law index
is somewhat flatter (see Table 4). The $\chi^2$ values are artificially 
high as a consequence of a known problem with the response matrix near
2~keV; the rest of the spectrum matches the model well.
Interestingly, the ionization
parameter $\tau = n_et$ is 2.7 times higher for the bright clumps;
taking the mean time since the emitting gas was shocked as constant
over the face of the remnant, this implies that the clump density $n_e
= 2.7$ times the inter-clump density. The emissivity of the plasma should
scale as $n_e^2$. In fact, the mean surface brightness of the clumps is just 
$n_e^2 \sim 8$ times the mean surface brightness of the remaining diffuse
emission, lending some credence to the model fits.  The total luminosity
of the thermal component in the 0.5-10~keV band is nominally 
$1.8 \times 10^{37}$ erg s$^{-1}$; however, given the uncertain temperature
and large column density, this estimate could be high by a factor of
several. Nonetheless, it one of the most luminous shell remnants in the Galaxy. 
Possible origins for the diffuse power-law component are discussed below.

\section{Discussion}

\subsection{The distance to \kes}

Becker and Helfand (1984) presented a 21 cm hydrogen absorption
spectrum for \kes\ which shows clear evidence of absorption at
negative velocities, indicating a location beyond the solar circle; we
adopt their distance of 19 kpc. Assuming a hydrogen spin temperature
of 100 K, these measurements indicate a neutral atomic hydrogen column
density to the source of $\sim2 \times 10^{22}$ cm$^{-2}$. Since
roughly half the hydrogen along any line of sight through the Galaxy
is either in ionized or molecular form, we should expect an X-ray
absorption column density of $\sim 4 \times 10^{22}$ cm$^{-2}$, in
excellent agreement with that 
derived from our fit to the pulsar wind nebula spectrum ($3.96 \times
10^{22}$ cm$^{-2}$). The 
lower $N_H$ values found in the various thermal fits reported in Table
2 and 3 are implausible; the discrepancy is probably due
to inadequate plasma models. We thus use the wind nebula value in
assessing the intrinsic luminosities of all components of the remnant.

\subsection{The X-ray luminosity of the pulsar and its PWN}

The nonthermal luminosity from the pulsar is $4.1 \times 10^{35}$ erg
s$^{-1}$ in the $0.5-10$ keV band, and the luminosity of its wind
nebula\footnote{We note here that there is an error in the X-ray luminosities 
for this source quoted in Blanton and Helfand (1996). The observed, rather
than intrinsic, X-ray fluxes were used. The correct values from the {\it ASCA} analysis ($L_x(0.5-8.0~\rm{keV})_{core} = 2.5 \times 10^{36}$ erg s$^{-1}$ and
$L_x(0.5-8.0~\rm{keV})_{shell} = 1.1 \times 10^{37}$ erg s$^{-1}$) agree
well with the values derived here.}
is $1.7 \times 10^{36}$ erg s$^{-1}$.  Both values
are second only to the Crab Nebula and its pulsar among known
Galactic objects. However, the efficiencies with which this pulsar
converts its rotational kinetic energy to X-rays in its magnetosphere
and in its surrounding nebula are both  greater than the
Crab. The pulsar value of $L_{x-pulsar}/\dot E\sim 1.6\%$ is more than
six times this value for the Crab, although it is very similar to the
value for PSR~J0540-693, the 50 ms pulsar in the Large Magellanic
Cloud. The value of $L_{x-nebula}/\dot E \sim 6.5\%$ is the highest
known -- in this case, comparable to that for the Crab, but several times higher
than that for PSR~J0540-693.

The similarity of the pulsars and their wind nebulae in \kes\ and
PSR~0540-693 is noteworthy. Morphologically, they are two of the best
representatives of the composite remnant class, with prominent radio
and X-ray shells and bright pulsar-driven synchrotron cores.
Their relatively high values of $L_{x-nebula}/\dot E$
could result from the confinement of the outflowing pulsar wind by the
evident shells. The characteristic age of PSR~0540-693 is roughly
twice that of \psr\ (1400 vs 700 yrs), although the detailed optical
study of Kirshner et al. (1989) found an expansion velocity for the
shell surrounding PSR~J0540-693 of $\sim 3000$ km s$^{-1}$ 
and a dynamical age of $\sim 760$
yrs. There are several striking differences between these two systems
however: their spin periods, period derivatives (and, thus, inferred
magnetic field strengths), and their shell diameters. It is not
implausible that very different pulsar birth parameters have led to
the difference in the remnant sizes.

\subsection{The large shell diameter of \kes}

The radius of \kes, 9.7 pc, is enormous for the young characteristic
age of its pulsar, implying a {\it mean} expansion velocity for the
remnant of over 13,000 km s$^{-1}$. For even a modest $5 M_{\odot}$ of
material ejected into a vacuum, the kinetic energy required is nearly
10$^{52}$ erg. For 10 $M_{\odot}$ of ejecta
going off into a medium with a mean density of 0.5 cm$^{-3}$, the
required kinetic energy reaches nearly 10$^{53}$ erg, comparable to
the entire gravitational binding energy of the neutron star whose
formation produced the explosion. The shell X-ray luminosity derived in
\S5 suggests a density closer to 1 cm$^{-3}$. Such kinetic energy values are
both unprecedented and highly implausible.

There are several possible ways to reduce the required explosion
energy. A smaller distance for the remnant would reduce the inferred
velocity, albeit only as $d^{0.5}$. The HI absorption spectrum of
Becker and Helfand (1984) leaves little possibility of reducing the
distance by more than $\sim 30\%$, in that all velocities are seen in
absorption to the solar circle and beyond.  Alternatively, the remnant
could be considerably older than the pulsar's characteristic age
$\tau_c = P/2\dot P$. This would run counter to the situation for
other remnants. For the Crab, $\tau_c = 1240$ yrs, while its true age
$t = 945$ yrs, for PSR~0540-693 $\tau_c = 1672$ yrs (Zhang et al. 2001)
while the dynamical age estimate is 760 yrs, and for PSR~J1811.5-1926
in G11.2-0.3, $\tau_c = 24,000$ yrs, while its association with SN
386AD means $t=1616$ yrs (Kaspi et al. 2001); a plausible explanation
in each of these cases is that the pulsar's initial period was not
negligible compared to its present spin rate (an assumption of the
approximation for $\tau_c$ defined as above).

Indeed, for a pulsar braking index $n= - \nu \ddot \nu / \dot \nu^2
= 3$, expected for braking by pure magnetic dipole radiation, the
characteristic age is {\it always} an upper limit to the true age. It is
possible for the true age to exceed $\tau_c$, however, if $n<3$. But
for the range of most measured indices ($\sim 1.4<n<2.9$), $t\le 5\tau_c$, and
is equal only for a very high initial spin rate, $P_o$. The recent results
of Mereghetti et al. (2002) suggest a value for n in the range
$1.86<n<2.48$ depending on the glitch history of the star. For high initial 
spin rates, this yields ages in the range $980<t<1770$ years.
If we push all the
numbers in one direction -- reduce the distance to 15 kpc, posit only
5 $M_{\odot}$ of ejecta, require an ambient density of $n = 0.1$
cm$^{-3}$, and double the age, thus requiring a small $P_o$ -- the energy
needed {\it just} to
explain the observed kinetic energy is $2 \times 10^{51}$ ergs, at the
upper end of the inferred distribution of supernova explosion
energies.

A consequence of requiring a fast initial spin rate for
\psr\ is that, given its enormous magnetic field of $5
\times 10^{13}$ G, the amount of rotational kinetic energy dumped into
the remnant shortly after the explosion will exceed the fraction of
the explosion energy itself that couples to the surrounding matter
(typically $\sim 1\%$). For $P_0 = 0.01P \sim 3$~ms, $E_{rot} = 2.2
\times 10^{51}$ ergs. As Blanton and Helfand (1996) showed, the
current total energy in radio-emitting relativistic particles and
magnetic field in the synchrotron nebula is only $\sim 10^{48}$
ergs. If \psr\ had been born with a spin period only a
factor of two or so below its current value, it could have just
supplied this inferred energy with none left over to help power the
expansion of the remnant. However, if it were born spinning a hundred
times faster, the current synchrotron nebula represents only $\sim
10^{-4}$ of its total energy output, with the remainder transferred to
the kinetic energy of the shell.

\subsection{The hard component to the shell emission}

As we showed in \S4, the X-ray emission from outside the clearly
demarcated pulsar wind nebula includes a hard spectral component that
is inconsistent with an extension of the thermal emission that
accounts for the bulk of the shell's X-ray luminosity. Both a very
high-temperature plasma ($T\sim 20$ keV)
and a power-law can satisfactorily fit the high-energy spectrum. We
explore here the possible origins for this emission.

There are two possible sources of relativistic electrons which could contribute
to the power-law emission: leakage of pulsar wind nebula
particles into the shell cavity, and particles accelerated at the
blastwave shock. The first explanation can be shown to be implausible.

There is no strong gradient in the amount of high energy emission with
radius; in particular, 5 keV photons in excess of those expected from
the low-temperature thermal plasma are found to the edge of the
remnant shell.  The particle energy required for 5 keV radiation $E
\sim 2.7 \times 10^4$ GeV. The synchrotron lifetime of
such a particle must, at minimum, exceed the light travel time from
the pulsar to the rim of the remnant: $\tau_{synch}>c/r\approx10^9$~s
or 32 years. This sets a constraint on the strength of the magnetic
field in the remnant $B<422~{\rm s}~\tau_{synch}^{-1}
E_{GeV}^{-1}\approx 4\mu$G -- comparable to the interstellar
field. By contrast, the equipartition field in the pulsar wind nebula is
$300\mu$G; furthermore, propagation of the particles from
the wind nebula boundary to the remnant's rim at $c$ is highly
unrealistic, lowering further the allowed field strength within the
shell. Thus, it seems highly unlikely that the bulk of the high energy
diffuse emission in the remnant is from pulsar-injected particles.

In recent years, synchrotron X-rays from particles accelerated at the outward
moving supernova shock have been discovered in several remnants (Allen,
Gotthelf \& Petre 2000 and references therein; Slane et al. 1999; Slane
et al. 2001). The observed power law spectral indices are quite steep,
ranging from $\Gamma = 2.4$ in G347.3-0.5 (Slane et al. 1999) to $\Gamma = 3.3$
in RCW 86 (although it should be noted that the XTE results of Allen et al.
(2000) on young remnants uses a harder energy band -- and yields steeper 
slopes -- than are obtained for the two ASCA synchrotron shells which
are fitted in a softer band). Contributions from nonthermal emission to the
total X-ray flux in the $1-10$ keV band range from a few percent to $>90\%$.

The fitted power law component in the shell of \kes\ has a photon
index on the flat end of this distribution (although it is not very 
well-constrained -- see Table 4) and an implied luminosity of
$\sim 3.2 \times 10^{35}$ erg s$^{-1}$. Given the high shell expansion
velocity, it is not implausible that some of this emission is indeed
direct synchrotron radiation from high energy electrons. However, when
the effect of dust scattering from the PWN is included (see below), the inferred
spectral index is flatter than that seen in any other remnant.

One component that must be present at some level arises from the
dust-scattering halo of the central nebula plus pulsar (e.g., Mauche
and Gorenstein 1986).  A substantial literature exists on this topic,
although it has been problematic to derive quantitative constraints on
the relevant parameters -- grain size distribution, composition,
internal structure, and distribution along the line of sight -- from
data of limited spatial resolution. The recent \chandra\ observation of the
Galactic X-ray binary GX 13+1 by Smith, Edgar, and Shafer (2002) shows
simultaneously the lack of agreement with previous models as well as
the sensitivity of such observations to various instrumental effects.
Nonetheless, as it provides a direct empirical measurement of
dust-scattering effects at the highest available resolution, we use
these results to estimate the contamination of the \kes\ shell by the
dust scattering halo of the PWN.

Smith et al. provide radial profiles at three energies for the halo of
GX 13+1.  We have measured the mean surface brightness in each
$10\asec$ annulus from $50\asec$ to $100\asec$ (the outer limit of the
remnant shell) and calculated the fraction of scattered flux
detected. We have then extrapolated the observed surface brightness to
$25\asec$ (the inner boundary of the diffuse shell emission in
question) and obtained a crude estimate of the fraction of the
intensity scattered into the SNR shell region from the pulsar plus its
surrounding nebula (ignoring the effect of the nebula's extent). We
obtain values of $\sim 15\%$, $\sim 10\%$, and $\sim 8\%$ for the
$\Delta E = 100$ eV energy bands centered at 2.15, 2.95, and 3.75 keV,
respectively. As noted by Smith et al., the energy dependence $\sim
E^{-1}$ is shallower than the theoretically expected value of
$E^{-2}$. We then correct for the difference in column density between
GX 13+1 and \kes\ ($\sim 2.8 \times 10^{22}$ cm$^{-2}$ vs.  $\sim 4 \times
10^{22}$ cm$^{-2}$) by finding $\tau_{scat}$ from Figure 4 in Predehl
(1997) and using the simple relation for the fractional scattered
intensity $I_{frac} = (1 - e^{-\tau_{scat}})$; this increases the
estimates above by a factor of 1.2.

In the band $2-4$ keV, then, it is possible to explain most of the
putative power-law emission as a simple dust-scattering halo of the
PWN.  In the $1-2$ keV band, however, we would expect a contribution
from the halo that exceeds the total power law component. Two effects
are relevant here. Mathis and Lee (1991) show that multiple scattering
is expected to set in for $\tau_{scat}>1.3$ and broaden the halo; for
our column density, this limit is exceeded at 1.7 keV, although the
effect is not large for the $25\asec$ to $100\asec$ region of
interest here. Secondly, the thermal component of the fit absorbs a
large (and uncertain) fraction of the power in the $1-2$ keV band where
the Mg and Si lines are found. If some of the photons in this band are
in fact attributable to dust scattering, it might well raise the
equivalent widths of these lines and boost their sub-solar abundances
to (the expected) higher values. It could also help explain the
disturbing trend toward lower temperatures at larger shell radii noted
above.

In the high energy ($4-7$ keV) band, there appears to be some power
law contribution beyond the expected halo emission; whether this is
evidence of an unusually flat-spectrum synchrotron contribution from the
shell, a small thermal contribution from the fast blastwave shock, or further
unexpected scattered flux remains unclear.

A quantitative analysis of the spectrum including all of the effects
discussed above is beyond the scope (not to mention the calibration
uncertainties and photon counting statistics) of this paper. Less
distant remnant shells are better targets for attempts to untangle the
complicated physics of thermal and non-thermal X-ray
emission. However, the apparent importance of considering
dust-scattering in this case should serve as a cautionary tale for
observers using the new high-energy imaging capabilities to explore
absorbed remnants in the Galactic plane.

\subsection{Conclusions}

Our high-resolution imaging observations of \kes\ have revealed all the
components expected surrounding the site of recent stellar demise. The
young pulsar and its wind nebula are shown to be among the most efficient
known at turning rotational kinetic energy into X-ray emission; the
extraordinary magnetic field strength of the pulsar may be responsible
for this notoriety.  Thermal X-ray emission is seen from throughout
the remnant shell, although deriving constraints on the remnant's
evolutionary state and elemental abundances is compromised by the
presence of a significant diffuse nonthermal component which we
attribute largely to the dust-scattering halo of the pulsar and its
nebula. The presence of such halos needs to be considered when
deriving inferences concerning the presence of nonthermal components
in the spectra of distant shell-type and composite remnants.

\begin{acknowledgements}
{\noindent \bf Acknowledgments} --- D.J.H. acknowledges support from the
\chandra\ program through grant SAO G00-1130X, while E.V.G. is supported by NASA
LTSA grant NAG5-7935.
\end{acknowledgements}

\clearpage
\twocolumn

\begin{figure}
\epsscale{1.0}
\plotone{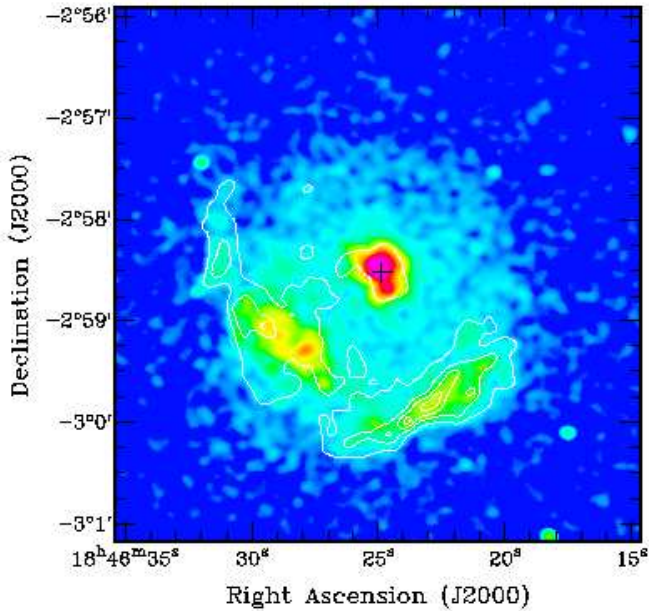}
\caption{ The X-ray view of \kes\ and its central pulsar wind nebula
in the $1.0-7.0$ keV energy band, as obtained with the ACIS instrument
on-board the \chandra\ Observatory. The pulsar \psr\ resides at the
center of wind nebula. The false color image is displayed with a logarithmic
intensity scale. Overlayed are the VLA 20~cm radio image
contours. The X-ray image has been smoothed with an elliptical Gaussian function
to match the radio image which has a synthesized beam of
$6\farcs2 \times 5\farcs4$. The similarity of the morphology in the
two wavelength regimes is apparent. The cross marks the position of the pulsar;
the point source is not apparent because the high-surface-brightness pixels have
been saturated to show the structure of the lower surface brightness emission.}
\end{figure}

\begin{figure}
\plottwo{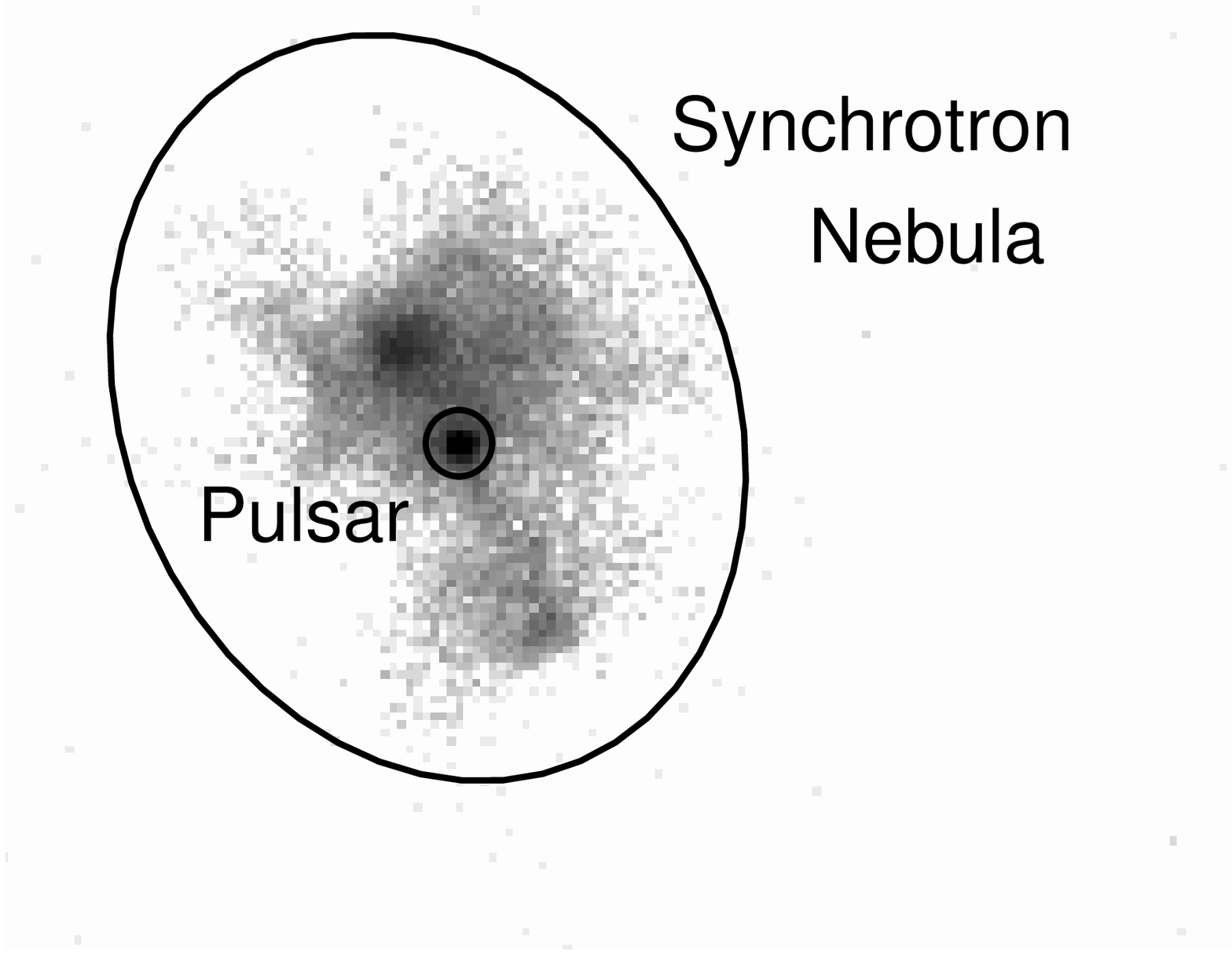}{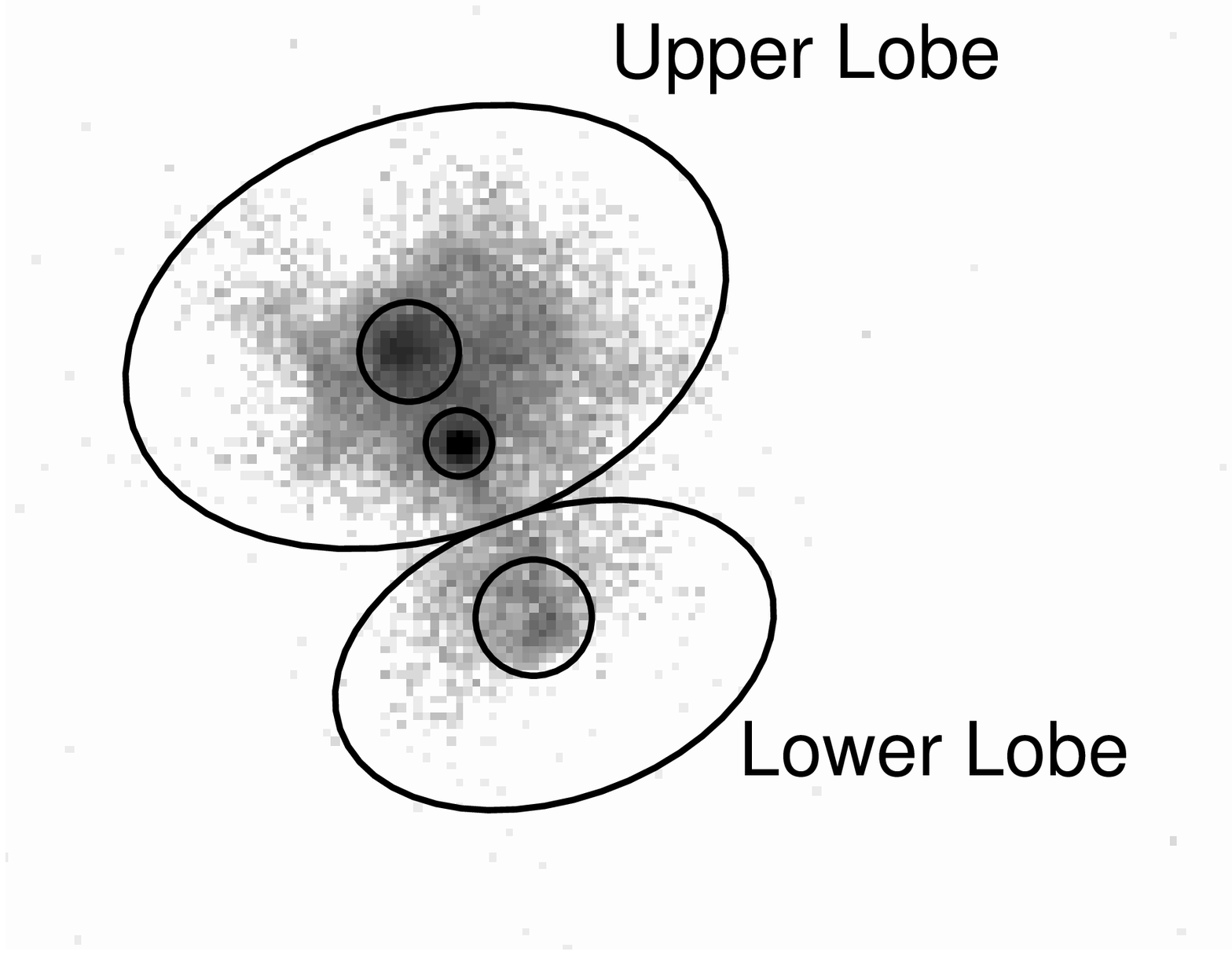}
\epsscale{1.0}
\bigskip
\caption{A blown-up X-ray view of the center of \kes\ showing the
pulsar wind nebula emission surrounding \psr, as obtained with the
ACIS-S3 instrument on-board the \chandra\ Observatory. {\bf (Left)} --
The central circle illustrates the aperture used to extract photons
from the pulsar, while the ellipse denotes the area (with the circle
excluded) used for extracting PWN photons.  {\bf (Right)} -- Same view
as the adjacent panel showing one of the sets of regions used in our
morphological analysis of the pulsar wind nebula (see \S 4.1). The
circles flanking the pulsar aperture isolate two bright spots. The
region of the pulsar aperture is excluded from the upper-lobe
area. The unbinned and unsmoothed image in both panels is displayed
with a logarithmic intensity scale.}
\end{figure}

\begin{figure}
\epsscale{1.0}
\plotone{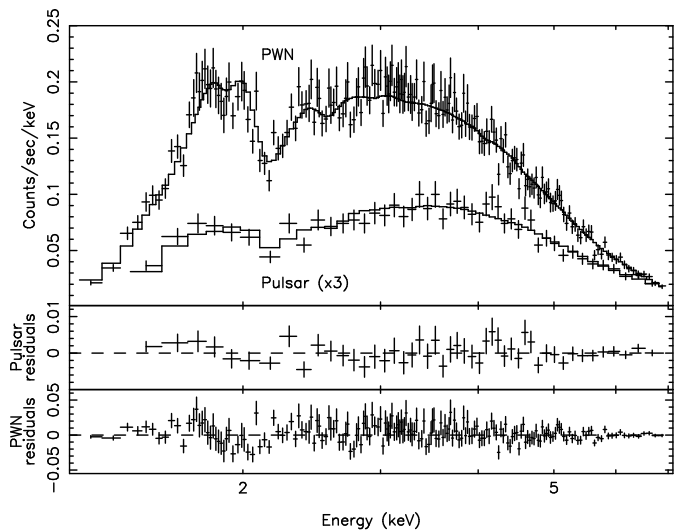}
\caption{Spectra and residuals for an absorbed power-law model fit 
to the data from \psr\ and the surrounding PWN. Spectra were extracted 
using the regions shown in Figure 2a. Both spectra are grouped to 
contain a minimum of $20$ counts per spectral bin.  The 
pulsar spectrum has been corrected for pile-up and background
contamination following the procedure described in \S 4.2}
\end{figure}

\begin{figure}
\epsscale{1.0}
\plotone{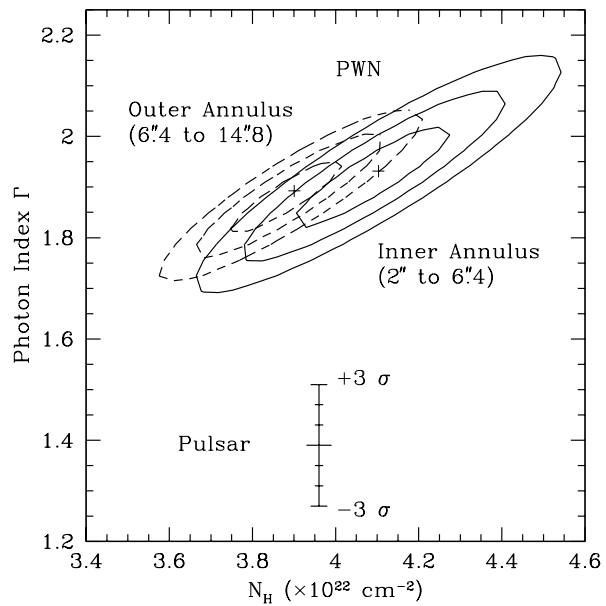}
\caption{Confidence contours for two interesting parameters for model
fits to inner and outer regions of the pulsar wind nebula using a simple 
absorbed power-law model. No evidence for a steepening of the power law
index with radius is apparent. The photon index for the pulsar (derived by
holding the $N_H$ value fixed at the best-fit nebular value) is significantly
flatter than that of the nebula after taking full account of source pileup (see
text for details).}
\end{figure}

\begin{figure}
\epsscale{1.0}
\plotone{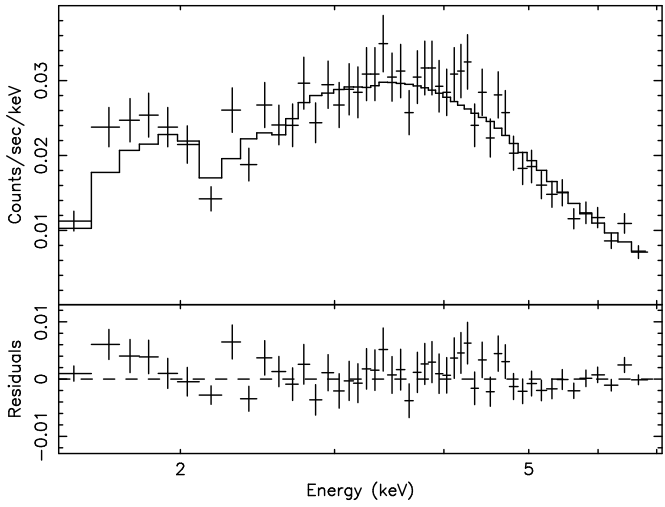}
\bigskip
\caption{ Best fit model for the spectrum of \psr\ using only the
2-7 keV band to search for a putatuive blackbody component. Note
the apparent excess in the
four lowest spectral channels. The model has been corrected for pile-up and
background contamination following the procedure described in \S 4.2}
\end{figure}

\begin{figure}
\epsscale{1.0}
\plotone{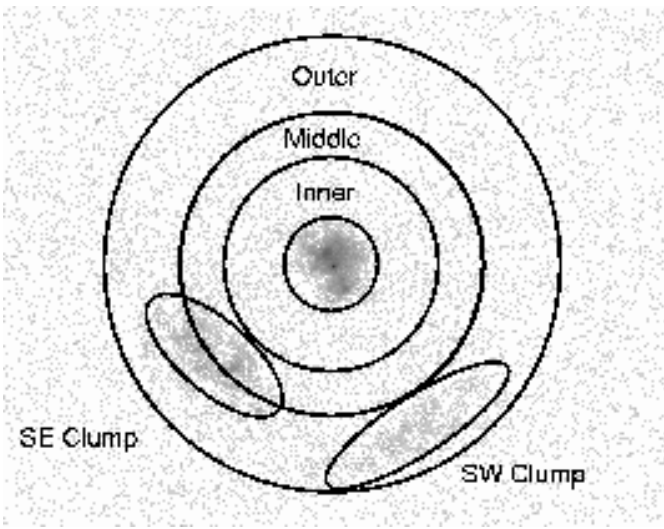}
\caption{ Extraction regions used for the spectral analysis of the
thermal emission from \kes\ showing the inner, middle, and outer
annuli and the two clumps.  The unbinned and 
unsmoothed image is displayed with logarithmic intensity}
\end{figure}

\begin{figure}
\epsscale{1.0}
\plotone{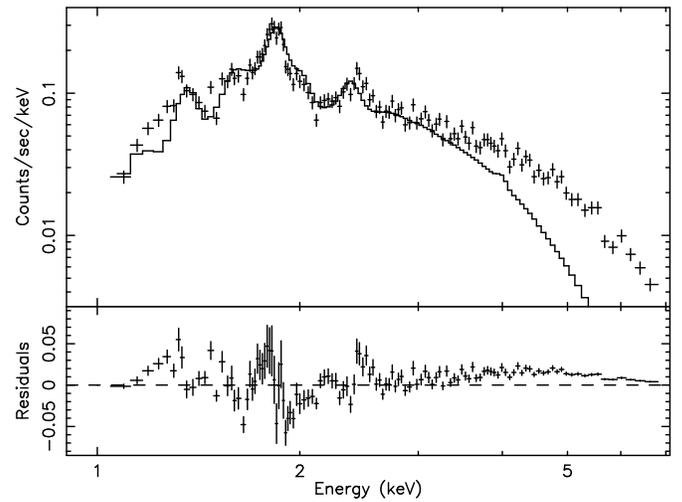}
\caption{ Spectrum and model fit to the diffuse emission from \kes.
Shown is the best fit non-equilibrium ionization collisional plasma
model.  The failure of this model to fit both the emission lines
and the hard continuum simultaneously properly is obvious. }
\end{figure}

\onecolumn

\begin{figure}
\epsscale{1.0}
\plottwo{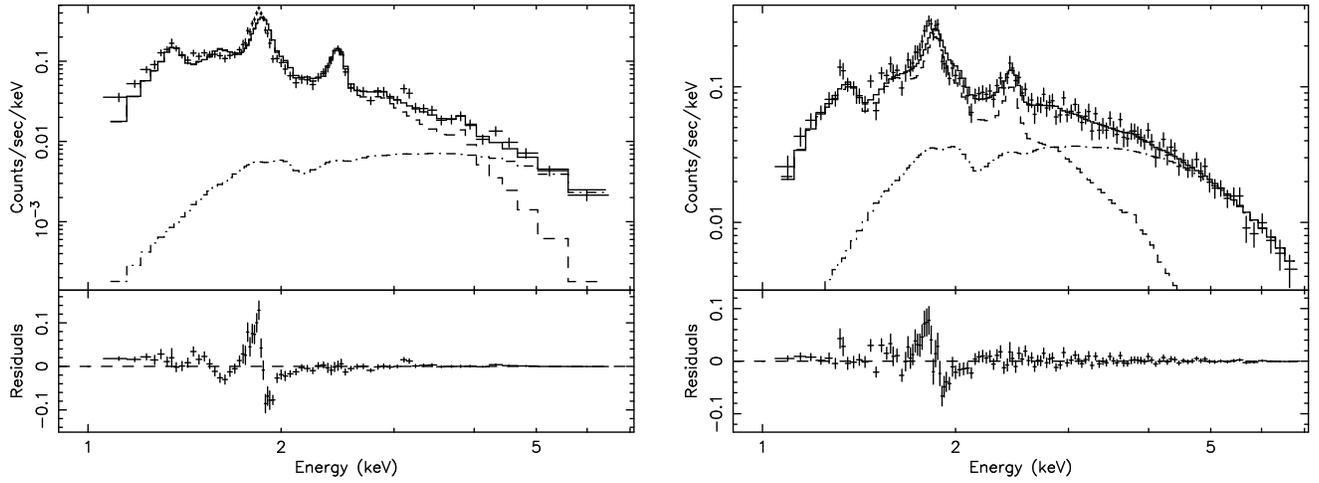}{f8b.eps}
\caption{Spectra from two regions of the thermal shell of \kes. These
spectra are fitted with a two component model including both a
non-equilibrium ionization collisional plasma model plus a power-law
component. {\bf Left} -- the spectra and best fit model for the south
eastern clump. {\bf Right} -- The same model as applied to the diffuse
emission region. In both cases the power-law component fits the higher end of
the spectrum very well, lowering the temperature of the thermal
component to a more reasonable value. The dominate feature around 2 keV 
in the residual plots are most likely due to a known response matrix
calibration issue.
}
\end{figure}

\onecolumn

\begin{deluxetable}{lcccc}
\tablewidth{0pt}
\tablecolumns{4}
\tablecaption{Spectal Fits to \psr\ and its PWN}
\tablehead{
\colhead{} & \colhead{$N_H$} & \colhead{Photon} &  \colhead{Flux\tablenotemark{a}}&  \colhead{} \\
\colhead{Region} & \colhead{($\times 10^{22}\ {\rm cm^{-2}}$)} & 
\colhead{Index} & \colhead{(ergs cm$^{-2}$ s$^{-1}$)} & \colhead{$\chi^{2}_\nu$ (DoF)}
}
\startdata
Pulsar\tablenotemark{b} & $3.96$ (fixed) & $1.39 \pm 0.04$ & $0.95\times 10^{-11}$ & 1.1 (425) \\
Pulsar Wind Nebula & $3.96 \pm 0.08$ & $1.92 \pm 0.04$ & $4.0\times 10^{-11}$  & 0.97 (256) \\
\ \ Inner Annulus & $4.08 \pm 0.13$ & $1.92 \pm 0.07$ & \dots & 1.11 (146) \\
\ \ Outer Annulus & $3.88 \pm 0.10$ & $1.88 \pm 0.05$ & \dots & 1.04 (259)\\
\enddata
\tablenotetext{a}{Flux quoted for the 0.5 -- 10 keV energy range.}
\tablenotetext{b}{This model requires additional processing due
to the effects of CCD pileup; the procedure used to produce this 
corrected spectrum is described in the text.}
\end{deluxetable}

\begin{deluxetable}{lccccl}
\small
\tablewidth{0pt}
\tablecolumns{5}
\tablecaption{Spectral Fits to the \kes\ Thermal Shell (Selected Regions)}
\tablehead{
\colhead{Parameter\tablenotemark{a}} & \colhead{SE Clump} & \colhead{SW Clump} & \colhead{Faint Diffuse} & \colhead{entire SNR}
 }
\startdata
$N_H$ ($10^{22}\ {\rm cm^{-2}}$) & $1.69 \pm 0.04$ & $1.74 \pm 0.08$ & $1.79 \pm 0.07$ & $1.74 \pm 0.03$ \\
$kT$ & $1.79 \pm 0.07$ & $2.44 \pm 0.22$ & $5.02 \pm 0.60$ & $2.99 \pm 0.12$ \\
Mg {\small XI}: \\
\ \ $E$ & $1.325 \pm 0.006$ & $1.330 \pm 0.008$ & $1.310 \pm 0.008$ & 
$1.321 \pm 0.004$ \\
\ \ $\sigma_E$ (keV)& 0 & 0 & 0 & 0 \\
Si {\small XIII}: \\
\ \ $E$ & $1.827 \pm 0.002$ & $1.826 \pm 0.004$ & $1.814 \pm 0.004$ & 
$1.825 \pm 0.002$  \\
\ \ $\sigma_E$ & 0 & 0  & $0.027 \pm 0.007$ & $0.010 \pm 0.006$  \\
S {\small XV}: \\
\ \ $E$ & $2.436 \pm 0.004$ & $2.430 \pm 0.008$ & $2.400 \pm 0.018$ & 
$2.427 \pm 0.005$  \\
\ \ $\sigma_E$ & 0 & $0.019 \pm 0.027$ & $0.146 \pm 0.021$ & $0.046
\pm 0.008$ \\
$\chi^2_\nu$ (DoF) & 1.22 (60) & 1.77 (58) &  1.21 (113) & 1.37 (142) \\
\enddata
\tablenotetext{a}{All units of energy are in keV. Spectral fits in the $1-7$ keV energy range.}
\end{deluxetable}

\begin{deluxetable}{lcccl}
\small
\tablewidth{0pt}
\tablecolumns{5}
\tablecaption{Spectral fits to the \kes\ Thermal Shell (Radial Regions)}
\tablehead{
\colhead{Parameter\tablenotemark{a}} & \colhead{Inner} & \colhead{Middle} & \colhead{Outer}}
\startdata
$N_H$ &  $2.45 \pm 0.09$ & $2.30 \pm 0.11$ & $1.84 \pm 0.08$ \\
$kT$ &  $4.87 \pm 0.58$ & $4.23 \pm 0.63$ & $3.11 \pm 0.32$ \\
Mg {\small XI}: \\
\ \ $E$ & $1.226 \pm 0.029$ & $1.301 \pm 0.014$ & $1.316 \pm 0.007$ \\
\ \ $\sigma_E$ & 0 & 0 & 0 \\
Si {\small XIII}: \\
\ \ $E$ &  $1.800 \pm 0.009$ & $1.797 \pm 0.014$ & $1.826 \pm 0.005$ \\
\ \ $\sigma_E$ &  0 & $0.032 \pm 0.022$ & $0.032 \pm 0.010$ \\
S {\small XV}: \\
\ \ $E$ &  $2.474 \pm 0.033$ & $2.401 \pm 0.051$ & $2.409 \pm 0.016$ \\
\ \ $\sigma_E$ &  $0.074 \pm 0.057$ & $0.129 \pm 0.050$ & $0.077 \pm 0.024$ \\
$\chi^2_\nu$ (DoF) & 1.12 (31) & 1.10 (34) & 1.17 (59) \\
\enddata
\tablenotetext{a}{All units of energy are in keV. Spectral fits in the $1-7$ keV energy range.}
\end{deluxetable}

\begin{deluxetable}{lcc}
\tablewidth{0pt}
\tablecolumns{3}
\tablecaption{Spectral fits to the \kes\ Thermal Shell (w/ Hard Comp. Model)}
\tablehead{
\colhead{} & \colhead{Diffuse} & \colhead{Clump} \\
\colhead{Parameter\tablenotemark{a}} & \colhead{Emission} & \colhead{Emission} }
\startdata
$N_H$ ($\times 10^{22}\ {\rm cm^{-2}}$) & $3.96$ (fixed) & $3.96$ (fixed)\\
\multicolumn{3}{c}{\it Abund. Allowed to Varying in Fixed Ratios} \\
$\Gamma$ &  $2.71 \pm 0.04$ & $3.02 \pm 0.09$ \\
$kT$ & $0.39 \pm 0.01$ & $0.66 \pm 0.01$\\
Abund. ($Z/Z_{\sun}$) & $13.6 \pm 0.4$ & $2.2 \pm 0.3$ \\
$\tau$ ($\times 10^{11}\ {\rm s\ cm^{-3}}$) &  $4.31 \pm^{0.91}_{0.66}$ &
$2.33 \pm^{0.25}_{0.23}$ \\
$\chi^{2}_{\nu}$ (DoF) & 2.73 (119) & 4.60 (135) \\
\multicolumn{3}{c}{\it Abund. Allowed to Vary Individually\tablenotemark{b}}\\
$\Gamma$ &  $1.72 \pm 0.07$ & $1.34 \pm 0.14$ \\
$kT$ & $0.69 \pm 0.01$ & $0.68 \pm 0.01$ \\
Mg & $0.35 \pm 0.04$ & $1.03 \pm 0.05$ \\
Si & $0.25 \pm 0.02$ & $0.66 \pm 0.02$ \\
S & $0.49 \pm 0.06$ & $0.98 \pm 0.05$ \\
Ca &  $0.35 \pm 0.60$ & $1.90 \pm 0.69$ \\
$\tau$ ($\times 10^{11}\ {\rm s\ cm^{-3}}$) & $0.809 \pm 0.060$ & $2.20
\pm^{0.24}_{0.21}$ \\
Flux\tablenotemark{c} \ (ergs cm$^{-2}$ s$^{-1}$) & $2.54 \times 10^{-10}$ & $1.66 \times 10^{-10}$ \\
$\chi^{2}_{\nu}$  & 2.15 (216) & 3.81 (132)\\
\enddata
\tablenotetext{a}{All units of energy are in keV. Spectral fits in the $1-7$ keV energy range.}
\tablenotetext{b}{Only the elements whose emission lines are primarily
within the energy range of our data have been fit as free parameters; 
the abundances of the elements with lines outside of that range have been fixed 
to the solar values.}
\tablenotetext{c}{Flux quoted for the 0.5 -- 10 keV energy range.}
\end{deluxetable}

\end{document}